# A new look at the anthropogenic global warming consensus: an econometric forecast based on the ARIMA model of paleoclimate series.

**Gilmar Veriato Fluzer Santos**[1]†*, Lucas Gamalel Cordeiro[1]†, Claudio Antonio Rojo[1]†, and Edison Luiz Leismann[1]†.

[1]Western Paraná State University, Graduate Program in Management/Professional Master's Degree, Cascavel, 85819-110, Brazil.

*Gilmar Santos, mail to: gviriato@hotmail.com

## Abstract

This paper aims to project a climate change scenario using a stochastic paleotemperature time series model and compare it with the prevailing consensus. The ARIMA - Autoregressive Integrated Moving Average Process model was used for this purpose. The results show that the parameter estimates of the model were below what is established by the anthropogenic current and governmental organs, such as the IPCC (UN), considering a 100-year scenario, which suggests a period of temperature reduction and a probable cooling. Thus, we hope with this study to contribute to the discussion by adding a statistical element of paleoclimate in counterpoint to the current scientific consensus and place the debate in a long-term historical dimension, in line with other research already present in the literature

**Keywords:** Global warming. Paleoclimatology. Time series. Arima model. Climate scenarios.

## Introduction

The controversies over global warming and its effects on the economy and the environment are the subject of discussion and debate around the world and in some ways determine how governments and companies develop their policies and conduct their business.

The human action according to the followers of anthropogeny and other international bodies such as the IPCC (Intergovernmental Panel on Climate Change - UN), has been responsible for climate change and global warming (greenhouse effect). This is endorsed by most scientific publications by showing that more than 90% of studies on the subject say that the cause of global warming is anthropogenic, established as the "official version" by IPCC advocates (Salzer, Neske & Rojo, 2019; Cook et al. 2013; Bray, 2010; Anderegg et al., 2010; Oreskes, 2004).

Shwed and Bearman (2010) bring an important contribution in the strategy of assessing the state of scientific contestations on certain issues when the scientific community considers a proposition a fact and how the importance of internal dissent in the face of consensus diminishes.

The IPCC Working Group Chair, Jim Skea, states: "Limiting warming to 1.5°C is possible within the laws of chemistry and physics but doing so requires unprecedented changes" (IPCC Special Report, 2019). On the other hand, the defenders of the naturalistic cause present arguments that challenge these studies by claiming that anthropogenic global warming is theoretically fragile with calculated misinformation, and its historical sample of only 150 years would be insufficient to establish a consensus often supported by agnotology and metric uncertainties (Molion, 2008; Legates *et al*. 2015; Legates, Soon & Briggs, 2013; Reinsinger *et al*. 2010).

What is noticeable is that the more research explores the past, the more the anthropogenic thesis is weakened, as demonstrated by Davis (2017) and Harde (2019) by finding that changes in the atmospheric $CO_2$ concentration did not cause changes in ancient climate temperature and climate change is not related to the carbon cycle, but rather to native impacts. Easterbrook (2016), in his evidence-based book brought data opposing $CO_2$ emissions as the primary source of global warming, the thesis of which has been captured by politics and dubious computer modeling.

Other anthropogenic studies ignore paleoclimatology as a relevant factor in research or have it as a factor of uncertainty, such as that of Haustein *et al*. (2017), Cook e*t al*. (2013), Mitchell *et al*. (2017), Medhaug e*t al*. (2017), in addition to those at the genesis of the IPCC studies (Solomon *et al*. 2007). However, increasingly scientists are pointing to data which suggests that climate changes are a result of natural cycles, which have been occurring for thousands of years, as Easterbrook (2016) and the arguments present in Koonin's book (2021).

Thus, it is possible to identify a gap in this debate which is a broader time horizon research and give statistical predictability to climate change. This is the objective of this study, whose central theme is to establish a climate prediction scenario for the next 100 years, based on a 12,000-year paleotemperature series (Holocene Period), plus the uncertainties that the data used predict., plus the uncertainties that the data used predict. To do this, we adopted the Autoregressive Integrated Moving Average (ARIMA) model, also known as Box-Jenkins, whose database was taken from the article by Kaufman et al. (2020), who applied five statistical methods of thermal reconstruction to verify global mean surface temperature (GMST) to the present day, which served as the basis for this research.

Results generated indicated the fragility of the anthropogenic thesis, which showed significant divergence from the scenario projected by the IPCC, which in its latest report predicted an increase of more than 1.5°C in the planet's temperature by 2050 (IPCC, 2019).

Therefore, we sought to establish one more variable for the global warming issue, in order to innovate the discussion and enable a technically critical approach, with the intention of comparing it with the consensus that prevails today.

**Results**

The parameters used to reach at the results were the median and the 5th and 95th percentiles representing the estimate of uncertainties with 90% confidence, as the authors themselves indicate by recommending that

> "Future users of this reconstruction use the full ensemble when considering the plausible Holocene GMST evolution. By representing the multi-method reconstruction as a single time series, the median of the ensemble may be best along with the 90% range of the ensemble to represent uncertainty."

(Kaufman et al., 2020, p.04).

For building the results, the data were represented graphically and fed the software IBM - SPSS Statistics, v. 22, for processing the ARIMA methodology - Box Jenkins methodology and the corresponding outputs according to each step of the calculation. Figure 1 shows the evolution of the 12k median of the data set extracted from Kaufmann *et al*. (2020) on a 100-year scale, with milestone "0" being the year 2019 (p. 8) calculated from the different reconstruction methods.

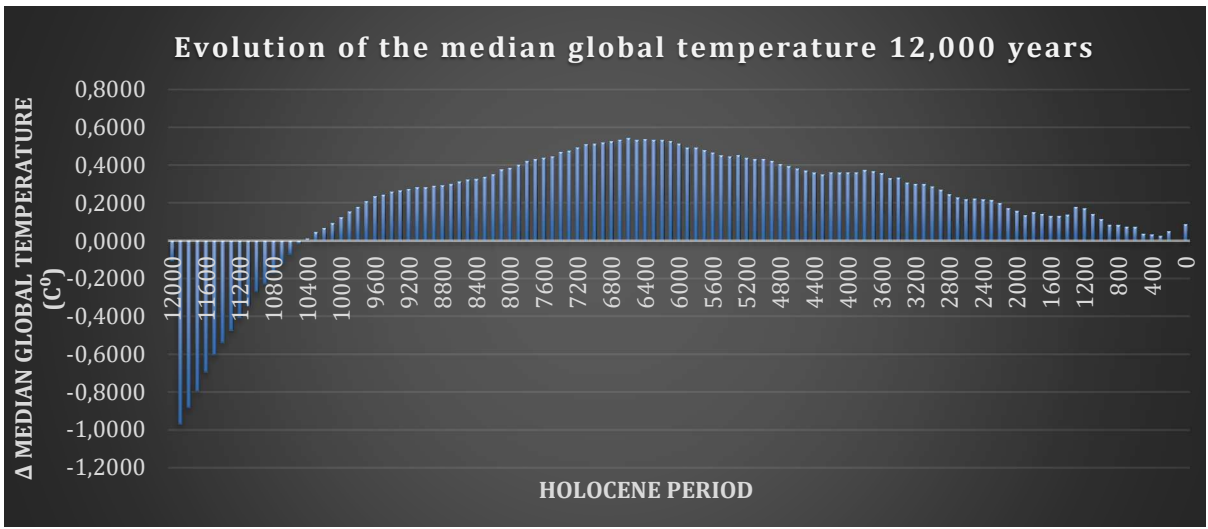

**Figure 1. Evolution of the Global Median 12k years temperature.**
Source: Author elaboration *(adapted from Kaufman et al. (2020 p. 06) from CSV file data at https://www.ncdc.noaa.gov/paleo/study/29712)*.

Figure 2 represents the 5th and 95th percentile range of the set bringing together the various sources of uncertainty, including proxy temperature, chronology, and methodological choices, as per Kaufman et al. (2020 p. 03).

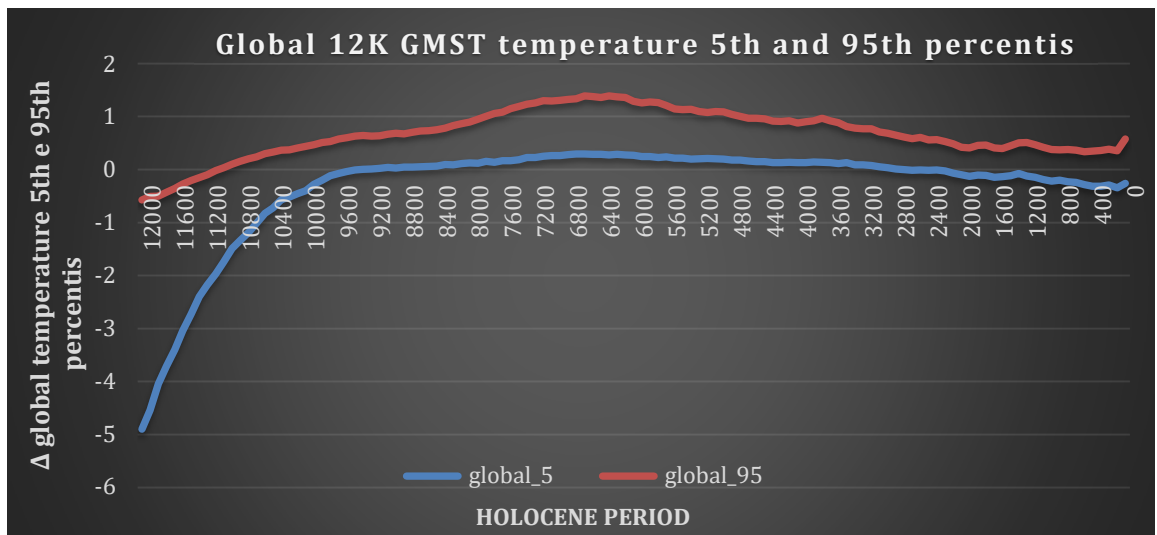

**Figure 2. Evolution of the parameters 5th and 95th global percentiles (uncertainties).**
Source: Author elaboration (*adapted from CSV file data - temp 12K all methods percentiles at https://www.ncdc.noaa.gov/paleo/study/29712)*.

The average temperature of the 1800-1900 period for each composite was used as the pre-industrial reference period defined by the authors as an anomaly of 0° C and which served as the reference for the IPCC (1850-1900). For this reason, it was removed from each member of the ensemble to avoid issuing individual records and different reconstructions (Kaufman et al, 2020).

Box-Jenkins ARIMA model's objective is to provide a valid basis for forecasting, after all tests, parameters, and diagnostics have been performed. The forecasts of the two-time series, median and uncertainties, were generated in the IBM - SPSS Statistics software, version 22, in a specific session for ARIMA modeling.

According to the model parameters, predictions for the median were expressed in the form of temperature estimates, for the next 100 years, represented by AR and MA. For statistical reliability purposes, the degree of significance (Marôco, 2018) of the parameters must be measured, being extremely significant in AR and very significant in MA, as described in figure 3.

**Arima model parameters**

| | | | | Estimate | SE | t | Sig. |
|---|---|---|---|---|---|---|---|
| MdTempGlob-Model_1 | **MdTempGlob** | No transformation | Constant | ,191 | ,129 | 1,489 | ,139 |
| | | | AR  Lag 1 | **,932** | ,032 | 28,799 | **,000** |
| | | | MA  Lag 1 | **-,266** | ,099 | -2,695 | **,008** |

**Figure 3:** 100-year scale temperature estimates of AR and MA parameters.
Source: Author elaboration with Software SPSS - Statistics v. 22.
(URL: https://www.ibm.com/support/pages/spss-statistics-220-available-download).

An important condition for model reliability is the residuals of the ACF and PACF correlations, *the white noise*. For the model to be validated as the most adequate, they should be concentrated around the mean, and the degree of significance is absolute (0 or close), thus represented in figure 4. (*Note: Retardo means Lag; "de resíduo" means of waste*)

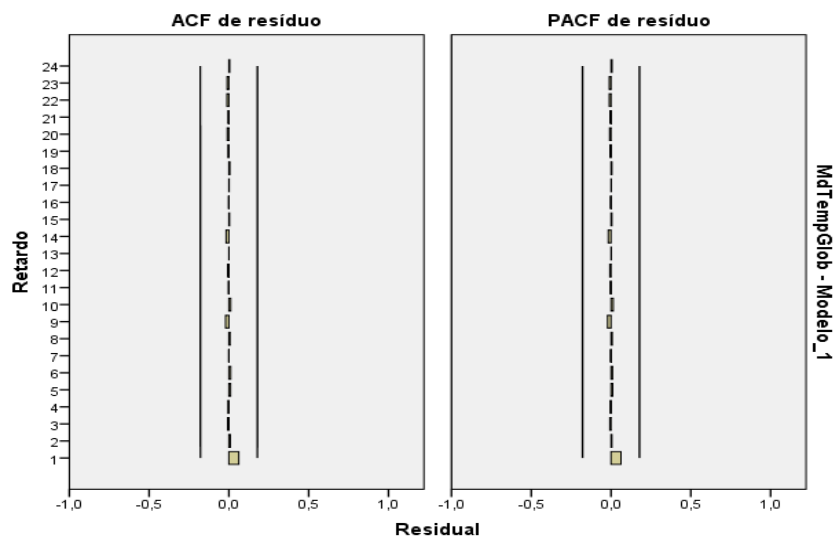

**Figure 4:** Residuals of the ACF and PACF correlograms (White noise).
Source: prepared by the author (SPSS - Statistics v. 22).

Thus, once stationarity is achieved (see p. 7-9), we can model it with an autoregressive process (AR), which we will represent by $Y_t$ the Median (Md) at period t (Holocene) as:

$$(Y_t - \delta) = \alpha^1(Y_t - \delta) + u_t \quad (1)$$

where δ is the mean of Y and $u_t$ is an uncorrelated random error with zero mean and constant variance $\alpha^2$ (this is white noise), then we will say that $Y_t$ follows a first-order stochastic autoregressive or AR process (1).

The AR process we have just discussed is not just a mechanism that may have generated Y. In this case, Y may evolve into a first order moving average process, or an MA (1). If we model Y in

this way:

$$Y_t = \mu + \beta_0 u_t + \beta_1 u_{t-1} \quad (2)$$

where μ is a constant and u, as before, is a white noise stochastic error term. Here Y at period t is equal to a constant plus a moving average of the current and past error terms. More generally, we can represent it like this

$$Y_t = \mu + \beta_0 u_t + \beta_1 u_{t\_1} + \beta_2 u_{t\_2} + \cdots + \beta q u_t - q \quad (3)$$

which is an MA(q) process. In short, a moving average process is just a linear combination of white noise error terms. In this case, most likely Y has characteristics of both AR and MA and is therefore ARMA. Then Yt follows an ARMA (1,1) process, and can be written as

$$Y_t = \theta + \alpha_1 Y_{t\_1} + \beta_0 u_t + \beta_1 u_{t\_1} \quad (4)$$

because there is an autoregressive term and a moving average term. In the Equation, ө represents a constant term. In general, in an ARMA (p, q) process, there will be p autoregressive terms and q moving average terms.

In the fit chart, shown in figure 5, it is observed that the two lines coincide, almost overlapping, indicating that this is the best of the models tested. The outliers present between 1 and 5 dates were kept in the setup since if we were to remove them, the series would not be robust. This guarantees its impartiality and uncertainty for future events (Stockinger & Dutter, 1987). *Note: observado means observed; ajuste means adjust; UCL: the upper control limit; LCL: the lower control limit.*

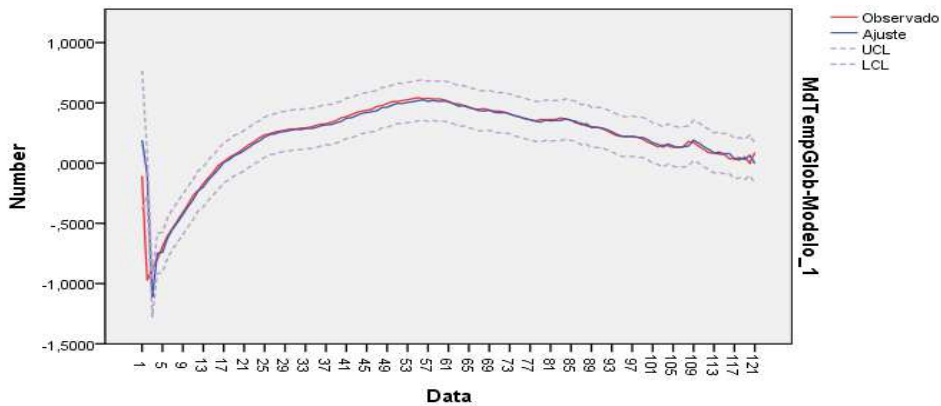

**Figure 5.** Graph of the adjusted 12k median series.
Source: elaborated by the author (SPSS- Statistics).

Regarding the uncertainty results 5th and 95th percentiles, the process follows the same model as the median, whose configuration is described in a supplementary file. The following parameters were generated, according to figure 6:

**ARIMA model parameters**

| | | | | Estimate | SE | t | Sig. |
|---|---|---|---|---|---|---|---|
| GLOBAL5-Model_1 | **GLOBAL5** | No transformation | Constant | -2,403 | 3,490 | -,688 | ,493 |
| | | | **AR** Lag 1 | ,999 | ,003 | 291,464 | **,000** |
| | | | **MA** Lag 1 | -,700 | ,069 | -10,111 | **,000** |
| GLOBAL95-Model_2 | **GLOBAL95** | No transformation | Constant | ,149 | ,684 | ,218 | ,828 |
| | | | AR Lag 1 | ,996 | ,006 | 179,006 | **,000** |
| | | | MA Lag 1 | -,382 | ,106 | -3,593 | **,000** |

**Figure 6.** Parameters of the 5th and 95th percentile temperatures (model uncertainty).
Source: Author elaboration (SPSS - Statistics v. 22).

We then have a set of six different extremely significant temperature results for the estimates of the two models: **0.932°C; -0.266°C** (fig. 3) and **0.999°C; -0.70°C; 0.996°C; -0.382°C** (fig. 6).

To fulfill the objective of this study, it is necessary that a standard measure be calculated and adopted as a reference. The median, extracted from the set of estimates of both models is the most appropriate statistical measure in this case, whose result was **0.333°C** (calculated from Microsoft Excel). It is evident, thus, a temperature below the 1.50% to 2.00% projected by the IPCC by the end of this millennium. It is evidenced, therefore, an average temperature well below the 1.50% to 2.00% projected by the IPCC by the end of this millennium. The results generated here indicate that, contrary to warming, the scenario outlined is that the world may experience a period of decreasing temperatures over the next hundred years, which could imply a cooling of global scope..

## Discussion

Given the results presented, one must ask why there is so much consensus around a scenario that as the evidence shows here, leaves much room for doubt? Another question that arises is why there is so much scientific unanimity around anthropogenic warming (97.2% according to Cook *et al.*, 2013), now called "climate change"?

It is understanding that if we compare recent temperatures to the distribution of global maximum temperatures during the Holocene, there was on average a 1°C increment over the pre-industrial period (1850-1900) and for most members of the ensemble, no 200-year interval during the series exceeded the warmth of the most recent decade (Kaufman et al., p. 5). We see, therefore, that the time horizon of the anthropogenic thesis is recent to the time of man's existence on earth (Holocene) and when compared to the results of this research, lacks substantiation if analyzed in the light of statistical science.

On the other hand, Kaufman et al., (2020), when relying on the IPCC projections, admit that temperatures for the rest of this century are likely to exceed 1°C if compared to those of the pre-industrial era (1800-1900), which they considered as an anomaly of 0°C. Although the authors claim that the Holocene GMST reconstruction is comparable with the IPCC long-term projections and those seen in the last decade, the results presented here show a different and antagonistic scenario if one considers a hundred-year scale and the historical temporality present in the statistical series.

Furthermore, in the graphical temporal observations of the studies by Kaufman et al. (2020, p.6, fig. 3), Davis (2017, p. 6, fig. 5) and Moberg et al. (2005, p. 3, fig. 2) we can see that there is significant climate variability every 2K years, so this casts doubt on establishing anthropogenicity as a criterion for the last 150 years. This is confirmed by the conclusion of Moberg et al. (2005) research when it states in its abstract and editorial summary that "The resulting model reconstruction supports the case that multicentennial natural variability has been larger than is commonly thought, and that considerable natural climate variation can be expected in future."

It is known that one of the villains of anthropogenic genesis, the greenhouse effect, was already unveiled in 1896 by Arrhenius as a natural phenomenon beneficial to the development of biological life on the earth's surface (troposphere) whose subsequent studies were duly confirmed (Miller & Spoolman, 2016). Therefore, reinventing this evidence is something that does not hold up considering the historical veracity of science, as the proponents of anthropogeny orthodoxly claim.

It does not exclude the impact that human action has brought to recent climate change, which might be important and timely, but seems to be insignificant in the face of the millennial variability of the climate, the size and complexity of the universe, and all the natural and astronomical phenomena that interact with the earth in the planetary system.

Lastly, it should be argued that the climate scenario predicted here is not enough to determine which are the true causes of recent climate change, whether natural or anthropogenic, since the two may be complementary, not divergent. For this, new studies on paleoclimate and its variability are needed to corroborate the estimates resulting from this research and to bring more evidence in the search for scientific truth.

It is unreasonable to subject the world and organizations to be hostages of a dubious thesis with all the consequences that this brings for the strategic planning of political and economic agents. In doing so, we may be condemning humanity to a climate catastrophism without any certainty to justify it.

## Data and methods

The data for this paper were collected from Kaufmann et al., (2020) unprecedented multi-method reconstruction research of mean land surface temperature (GMST) during the Holocene era (12,000 years) to the present day, "whose database is the most comprehensive global compilation of previously available published Holocene proxy temperature time series" (Kaufman et al., 2020, p. 01).

Extraction of the primary data from this study is available as individual CSV files and merged as a netCDF file at figshare 35 and at NOAA Palaeoclimatology 36 (https://www.ncdc.noaa.gov/paleo/study/29712). A CSV file with the multi-method joint median and 5th and 95th percentiles is also available in both data repositories. All were used as **input data** to compose the 12k time series of paleotemperatures in the two variables and fed into IBM SPSS-Statistics software (v. 22) for the calculation of parameters and estimates. The data generated for the development of this research are available in supplementary file.

## Stochastic Processes and the Stationarity Test

To introduce the development of the forecast, we justify graphically and mathematically the results that the SPSS software generated with their respective outputs, in the two variables of this study, the median and the uncertainty set. To better understand, we will use the graphs in this section and the mathematical formulation of their results as well as the structuring of the uncertainty set (same pattern) in a supplementary file.

First, it is necessary to apply two tests to verify the **stationarity** of the time series: **(1) graphical analysis** and **(2) the correlogram test**, since it is a condition for using the ARIMA (BJ) model.

By analyzing Figures 1, 2 and 5, we verify that the series **are not stationary,** that is, by establishing a mean line for the 12K global temperature median series (Figure 7) we verify that the data do not circulate around it and express a trend. *Note: número de sequência means sequence*

*number.*

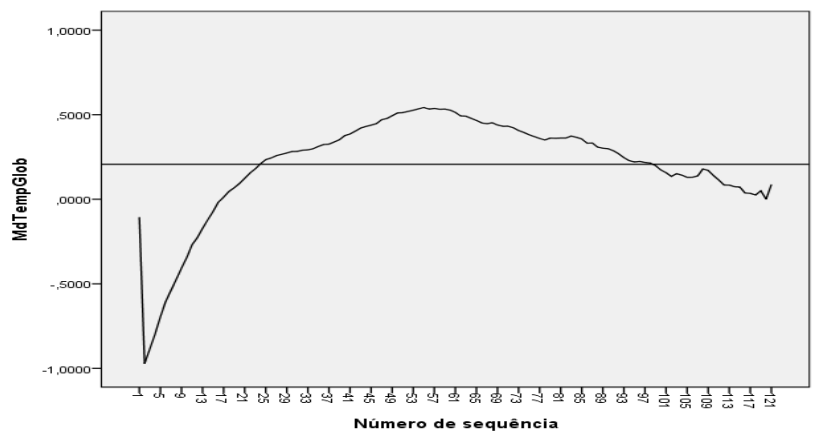

**Figure 7. Graphical test for stationarity.**

Next, we apply the correlation tests, also called "F" correlation function: ACF (automatic) and ACFP (partial), the next step to make the series stationary, as shown in figures 8 and 10, and their respective reports, represented by figure 9. *Note: coeficiente means coefficient. Número de retardo means Lag numbers.*

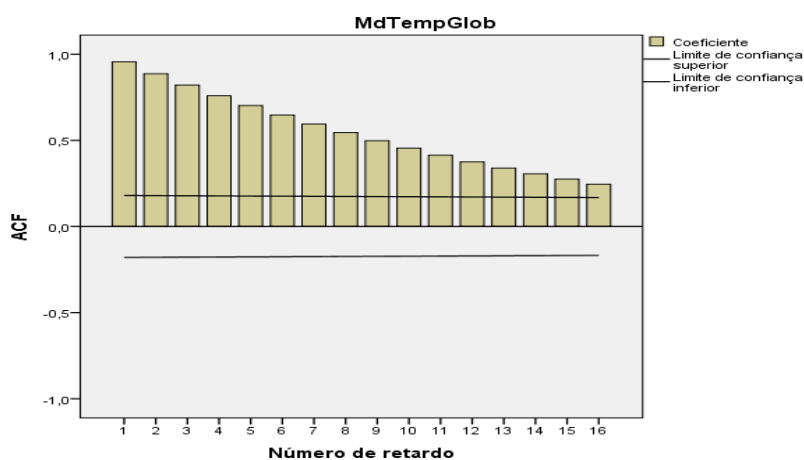

**Figure 8.** Graphical test of autocorrelation (automatic).

**Automatic correlations**

Series: MdTempGlob

| Lag | Autocorrelation | Standard Error[a] | Box-Ljung Statistics | | |
|---|---|---|---|---|---|
| | | | Value | df | Sig.[b] |
| 1 | ,956 | ,090 | 113,376 | 1 | ,000 |
| 2 | ,887 | ,089 | 211,775 | 2 | ,000 |
| 3 | ,821 | ,089 | 296,777 | 3 | ,000 |
| 4 | ,759 | ,089 | 370,121 | 4 | ,000 |
| 5 | ,702 | ,088 | 433,357 | 5 | ,000 |
| - | - | - | - | - | - |
| - | - | - | - | - | - |
| 16 | ,246 | 0,84 | 723,810 | 16 | ,000 |

a. The underlying process considered is independence (white noise).
b. Based on the asymptotic chi-square approximation.

**Figure 9.** LJung Box statistical report (Ho and H₁ hypotheses).

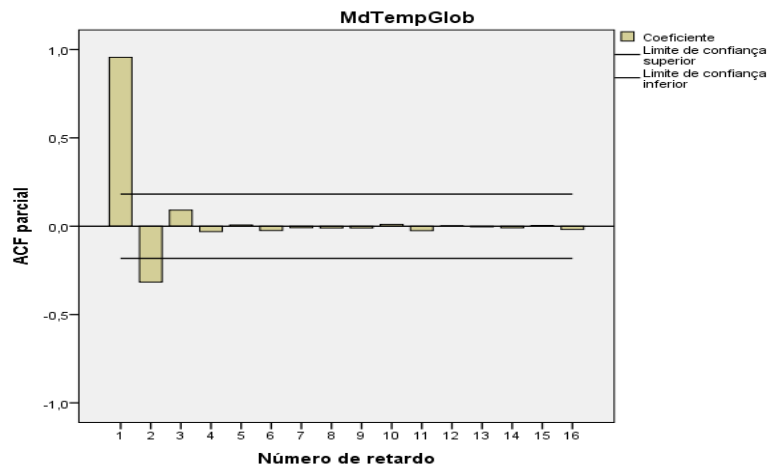

**Figure 10.** Graphical test of partial autocorrelation - PCA. Graphical and correlation analysis indicates that we have to normalize the series making it stationary**.** The process occurs with the choice of the first lag (lag), which exceeded the confidence interval in both tests and whose degree of graphical significance is higher, i.e., has the highest correlation and the lowest value according to the Ljung-Box statistic. The lag that meets these criteria, therefore, is number 1, highlighted in fig. 9.

From these results, we can graphically represent (figure 11) the **stationarity** adjusted, as a function of the first differentiation (lag 1):

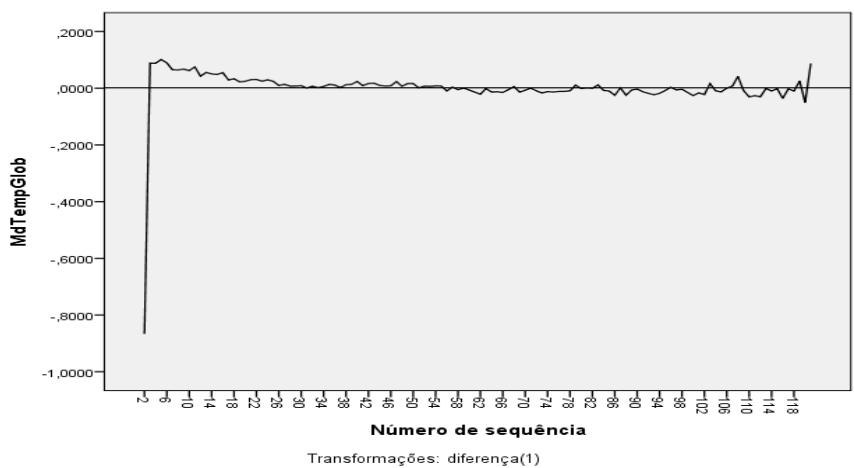

**Figure 11.** Adjusted stationarity as a function of lag 1. So, we can then replicate this modeling for the probabilistic analysis of the uncertainties, represented by the 5th and 95th percentiles, at a 90% confidence level, since it assumed the same stationarity criteria and tests (graph and correlogram) of the median. The graphical representation of the uncertainty set is described in the supplementary file.

### Applying the Box- Jenkins model

Box-Jenkins's method aims (figure 12) is to estimate a statistical model and interpret it according to the sample data. If this estimated model is used for forecasting, we should assume that its characteristics are constant over the period and particularly in future periods. A simple reason for requiring the stationary data is that any model that is inferred based on that data can be interpreted as stationary or stable and therefore provides a valid basis for prediction (Pokorny, 1987, Gujarati and Porter, 2011).

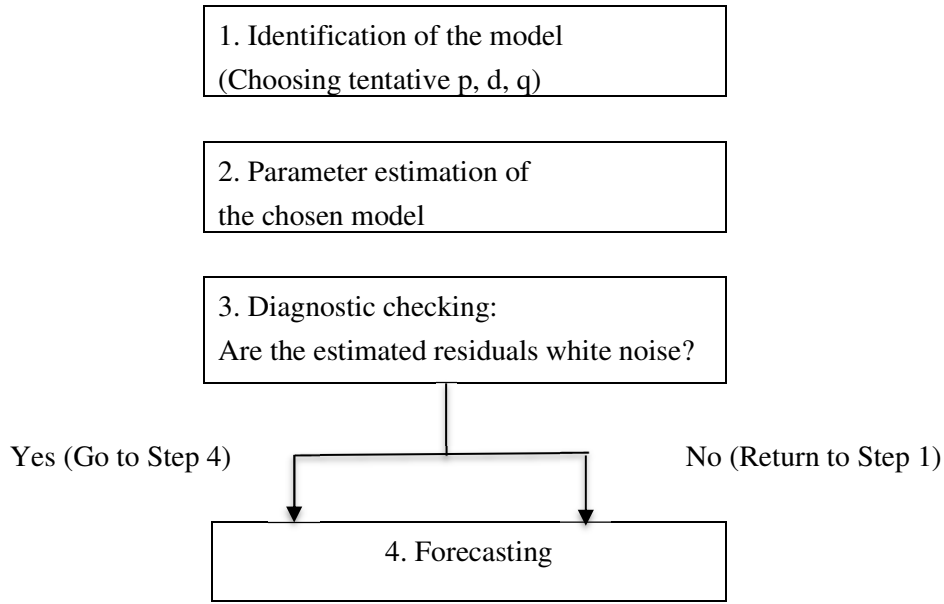

**Figure 12**. The Box–Jenkins methodology. About s**tep 4, Forecasting:** One of the reasons for the popularity of the ARIMA modeling is its success in forecasting. In many cases, the forecasts obtained by this method are more reliable than those obtained from the traditional econometric modeling, particularly for short-term forecasts. Of course, each case must be checked (Gujarati and Porter, 2011, p. 778).

We concluded that the MedTempGlobal (as described in the data/figures) time series model was not stationary and we had to normalize it, making it stationary with constant mean and variance and its covariance invariant over time. Therefore, it is an integrated time series, i.e., it combines the two autoregressive processes (AR and MA) in the same set.

An important point to note is that when using the Box- Jenkins methodology, we must have both a stationary time series and a time series that is stationary after one or more differentiations (Gujarati and Porter, 2011).

Then, we can state that if a time series is integrated of order **1**, therefore, it is **I (1)**, after differentiating it becomes **I (0)**, that is, stationary. In general, if a time series is **I** *(d)*, after differentiating it *d* times, we get an **I (0)** series.

If one has to differentiate a time series *d* times to make it stationary and apply the ARMA (p, q) model to it, one will say that the original time series is **ARIMA *(p, d, q)***, that is, it is a moving average **integrated autoregressive time series**, where p denotes the numbers of the autoregressive terms, *d* the number of times the series must be differentiated before it becomes stationary, and *q* the number of moving average terms.

We, therefore, have in this time series an ARIMA (1,0,1) model, as it was differentiated once (d = 1) before becoming stationary (of first difference), and can be modeled as an ARMA (1,1) process, as it has an AR term and an MA post stationarity.

Finally, it is important to emphasize that to optimize the results, it was necessary to run in the software SPSS - Statistics all the possible combinations of the ARIMA model *(p,d,q)* in the two parameters, to arrive at the statistically optimal model after the decomposition of the data and meeting the criteria of analysis and execution.

**Data Records**

The data that led to this research were reused by Kaufman et al., 2020, as already referenced in the text. After treatment, the data fed the IBM-SPSS Statistics v. 22, at https://www.ibm.com/docs/en/spss-statistics/SaaS?topic=reference-arima, for the development of this research and the generation of results. They can be found in the figshare repository: https://figshare.com/articles/dataset/ArimaMedTempGlobal_spv/14429006;https://figshare.com/articles/dataset/Spreadsheet_for_entering_and_processing_paleoclimate_data_and_graphs_with_the_results_of_the_model_/14429273;https://figshare.com/articles/dataset/Mathematical_and_operational_foundations_of_the_model_mediana_and_uncertainties_/14442701.

**Technical Validation**

All validations aiming to verify the technical quality and accuracy of the results were done in the ARIMA platform of SPSS-Statistics and are described in the body of the text and in the data repository. For space reasons, only the data from the model that satisfied the research methodology was sent, according to the foundations found in the specific literature.

**Acknowledgements**

The scientific foundation on which the results of this paper are based, without ideological or political bias, and the timing of the research data are important to consider. No less important is the legacy left by the authors cited and researched in this work, besides Kaufman et al. (2020): Routson et al. (2019), Cook et al. (2013), PAGES 2k Consortium, Marcott et al. (2013), Harde (2019), Box & Jenkins (1978), Gujarati & Porter (2011). Special gratitude and acknowledgement to professors

Claudio Antonio Rojo and Edison Luiz Leismann, who gave all the trust and support necessary to carry out this research.

**Author contributions statement**

G. V. F. S. and L.G.C designed the experiment (s), C. A. R. and E. L. L. conducted the experiments, E. L. L. and C.A. R. analyzed the results. All authors reviewed the manuscript.

**Competing interests and funding sources**

The authors declare no competing interests and have not received any specific grants from funding agencies in the public, commercial, or non-profit sectors to do this research.

**Additional information**                                                                                       .

**Correspondence** and requests for materials should be addressed to G.V. F. S.

**Author's information**

G.V.F.S: https://orcid.org/0000-0001-6216-2126 mail to: gviriato@hotmail.com

L.G.C: https://orcid.org/0000-0002-9043-7024

C. A. R: https://orcid.org/0000-0003-4484-9033

E. L. L: https://orcid.org/0000-0002-4112-8241